# Visuospatial short-term memory and dorsal visual gray matter volume

**Dennis Dimond**[a,b,c]**, Rebecca Perry**[b,d]**, Giuseppe Iaria**[b,e,f]**, Signe Bray**[b,c,d,g]

[a] Department of Neuroscience, Cumming School of Medicine, University of Calgary, Calgary, AB, Canada

[b] Alberta Children's Hospital Research Institute, University of Calgary, Calgary, AB, Canada.

[c] Child and Adolescent Imaging Research Program, University of Calgary, Calgary, AB, Canada.

[d] Department of Paediatrics, Cumming School of Medicine, University of Calgary, Calgary, AB, Canada.

[e] Department of Psychology, University of Calgary, Calgary, AB, Canada.

[f] Hotchkiss Brain Institute, Cumming School of Medicine, University of Calgary, Calgary, AB, Canada

[g] Department of Radiology, Cumming School of Medicine, University of Calgary, Calgary, AB, Canada

**Correspondence:** [*] **indicates corresponding author**

Dennis Dimond[*]; dennis.dimond@ucalgary.ca ; Alberta Children's Hospital, 2888 Shaganappi Trail NW, Calgary, Alberta, Canada, T2B 6A8.

## Abstract

Visual short-term memory (VSTM) is an important cognitive capacity that varies across the healthy adult population and is affected in several neurodevelopmental disorders. It has been suggested that neuroanatomy places limits on this capacity through a map architecture that creates competition for cortical space. This suggestion has been supported by the finding that primary visual (V1) gray matter volume (GMV) is positively associated with VSTM capacity. However, evidence from neurodevelopmental disorders suggests that the dorsal visual stream more broadly is vulnerable and atypical volumes of other map-containing regions may therefore play a role. For example, Turner syndrome is associated with concomitantly reduced volume of the right intraparietal sulcus (IPS) and deficits in VSTM. As posterior IPS regions (IPS0-2) contains topographic maps, together this suggests that posterior IPS volumes may also associate with VSTM. In this study, we assessed VSTM using two tasks, as well as a composite score, and used voxel-based morphometry of T1-weighted magnetic resonance images to assess GMV in V1 and right IPS0-2 in 32 healthy young adults (16 female). For comparison with previous work, we also assessed associations between VSTM and voxel-wise GMV on a whole-brain basis. We found that total brain volume (TBV) significantly correlated with VSTM, and that correlations between VSTM and regional GMV were substantially reduced in strength when controlling for TBV. In our whole-brain analysis, we found that VSTM was associated with GMV of clusters centered around the right putamen and left Rolandic operculum, though only when TBV was not controlled for. Our results suggest that VSTM ability is unlikely to be accounted for by the volume of an individual cortical region and may instead rely on distributed structural properties.

## 1. Introduction

Visual short-term memory (VSTM) is the ability to hold visual information in memory for several seconds (Baddeley, 2003). VSTM provides an essential gate between perception and cognition, allowing the retention of sensory information for higher cognitive processing (Baddeley, 2003). Although the precise limits on VSTM capacity are debated (Cowan, 2001)(and appended commentaries), it is clear that capacity varies between individuals (Conway et al., 2001; Vogel & Machizawa, 2004; Vogel et al., 2005).

VSTM changes across development (Park et al., 2002; Riggs et al., 2011) and is affected in many neurodevelopmental (Attout et al., 2017; Costanzo et al., 2013) and mental health conditions (Girard et al., 2018). Understanding the neuro-anatomical basis of VSTM has the potential to inform our basic understanding of cognitive capacity and has clinical relevance to disorders affecting VSTM and the potential for remediation. The map theory of VSTM capacity postulates that topographic maps found in early sensory (Sereno et al., 1995) and higher cognitive (Silver & Kastner, 2009) regions may place limits on cognitive capacity, as items must compete for actual, 'bounded', space (Franconeri et al., 2013). According to this model, one would predict a linear relationship between VSTM capacity and gray matter volume (GMV) of specific cortical regions. Such a relationship has indeed been reported by two recent neuroimaging studies; one reporting a positive association between VSTM and GMV of the primary visual cortex (V1) (Bergmann et al., 2016), the other reporting findings in the left inferior parietal lobule (Konstantinou et al., 2017). These findings provide support for the map-based theory of VSTM capacity and raise the possibility that other cortical areas might also play a role in limiting VSTM.

The intraparietal sulcus (IPS) is a cortical region that plays an important role in visuospatial attention and memory (Corbetta & Shulman, 2002). The IPS is topographically organized into 6 distinct regions, each containing a  topographic 'map' of the contralateral visual field (Silver & Kastner, 2009). The posterior IPS (IPS regions 0-2)  makes white matter connections to early visual areas such as V1 (Bray et al., 2013; Greenberg et al., 2012), and is believed to send top-down signals to influence sensory salience in the visual cortex (Lauritzen, D'Esposito, Heeger, & Silver, 2009). These brain regions (V1 and IPS0-2) are part of the dorsal visual stream, which has been suggested to be particularly vulnerable to neurodevelopmental abnormalities (Braddick et al., 2003). Deficits in dorsal stream function are present in several neurodevelopmental disorders (Atkinson et al., 2003) including patients with Williams and Turner syndrome, who also show deficits in VSTM (Attout et al., 2017; Cornoldi et al., 2001; Costanzo et al., 2013). In the case of Turner syndrome, VSTM deficits are concurrent with reduced GMV in posterior parietal and occipital regions (Zhao & Gong, 2017), as well as abnormal shape of the right posterior IPS (Molko, Cachia, Riviere et al., 2003). Taken together, the known involvement of IPS0-2 and V1 in visuospatial processing, along with concurrent findings of VSTM impairment and right posterior IPS and visual cortex abnormalities in TS, suggests that these regions may play a central role in limiting VSTM capacity and related deficits.

Taken together, a map-based theory of VSTM capacity, in combination with findings in neurodevelopmental populations, suggests that GMV of the right posterior IPS regions 0-2, in addition to early visual regions such as V1, may place constraints on VSTM capacity. In the present study we used voxel-based morphometry of T1-weighted magnetic resonance images to extract GMVs of V1 and IPS0-2 regions of interest (ROIs) using a probabilistic atlas (Wang et al., 2014), in 32 healthy young adults. We characterized VSTM using two assessments and correlated volumetric with cognitive measures. For comparison with previous findings, we also assessed associations between voxel-wise GMV and VSTM on a whole-brain basis. We found that total brain volume predicted VSTM, and individual ROIs did not add predictive value. At the whole-brain level, VSTM associated with GMV in the left parietal/temporal lobes and right putamen, though only when total brain volume was not accounted for. Our results suggest that VSTM is unlikely to rely on individual brain regions and may be more accurately associated with distributed representations.

## 2. Methods

### 2.1 Participants

All participants provided informed written consent, and the study was approved by the Conjoint Health Research Ethics Board at the University of Calgary. Inclusion criteria for the study were having normal or normal-corrected vision and right handedness. Participants were recruited from the University of Calgary student community and consisted of 32 young adults (aged 18.1 – 36.1, mean = 22.2, SD = 3.69, M/F = 16/16) having no history of neurological or psychiatric conditions. Participants received financial compensation for participating.

### 2.2 VSTM measurement

VSTM was assessed using a set of computer-based tasks programmed in the PsychToolbox (http://psychtoolbox.org/) in MATLAB (The Mathworks Inc., Natick, MA, USA). VSTM was assessed in a session that preceded MRI acquisition by several hours up to one month. Two assessments of VSTM were administered; the dot memory task, an assessment of location VSTM, and the modified span board task, which assesses location VSTM, but also includes a cognitive control component as it requires remembering both the location and order in which stimuli were presented.

In the dot memory task (Miyake et al., 2001), participants viewed a 5x5 grid and were asked to remember the location of items that flashed in the grid. Throughout the task a grayscale background was shown. At the start of each trial they were shown black gridlines for 1s followed by a set of black circles that appeared in specific locations in the grid for 750ms. The circle diameter was equivalent to the height and width of the grid cells. The dots disappeared, and participants were instructed to click in the grid cells to indicate where the dots were shown. There were no time limits within trials. The number of locations to remember increased from 3-7 across a total of 25 trials, with 5 trials for each number of locations to remember. Between each level there was a 2s pause during which a fixation cross was shown. The task was scored as the total number of correctly remembered locations out of a maximum score of 125. Participants started with 3 practice trials (3, 4 and 5 dots shown) to familiarize them with the task before the scored trials started. No performance criteria were imposed for the practice trials.

In the modified span board task (Westerberg et al., 2004), participants were asked to recall the spatial position of stimuli and also the order. A grayscale background was shown throughout this task. Each trial began with the presentation of a number in the center of the screen, indicating the number of squares that would be illuminated on the trial. This number was presented for 1s during the intertrial interval. On each trial, participants were shown a set of 9 small black squares irregularly spaced on the screen for 500ms. A set of squares were illuminated in white for 650ms one after the other with no temporal gap. Participants were then asked to click the squares in the order they were illuminated. There were no time limits within trials. Once clicked, the squares remained illuminated. The number of illuminated squares increased from 4-8 across a total of 25 trials, with 5 trials for each number of illuminated squares. The task was scored as the total number of squares recalled in the correct order, out of a maximum of 150. Participants completed two practice trials to become familiar with the task. No performance criteria were used for the practice trials.

Since combining tasks assessing the same cognitive construct can reduce measurement error and potentially provide a more accurate assessment, we also calculated a composite VSTM score by averaging Z-scores of the dot memory and span board tasks.

## 2.3 MRI data collection and processing

T1-weighted anatomical MR images (3D SPGR, 180 slices, FOV = 25.6cm, voxel size = 1mm isotropic, flip angle = 12) were collected on a 3T GE Signa scanner with an 8-channel head coil at the Seaman Family MR Research Center at the University of Calgary. We assessed gray matter volume (GMV) with voxel-based morphometry (Ashburner & Friston, 2000) in SPM12, using standard processing procedures. In brief, T1-weighted images were segmented, images were co-registered to a study specific template generated using the Dartel toolbox (Ashburner, 2007), and subsequently normalized to standardized Montreal Neurological Institute (MNI) template space (Mazziotta, Toga, Evans, Fox, & Lancaster, 1995). Motivated by concurrent findings of impaired VSTM and structural abnormalities in Turner syndrome, we took an ROI-based approach to assess GMV of the bilateral primary visual cortex, and right posterior IPS regions 0-2. GMV was averaged within a set of 4 ROIs; V1 (left and right combined), right IPS0, right IPS1 and right IPS2. All ROIs were defined using a probabilistic atlas (Wang et al., 2014).

## 2.4 ROI Statistical analyses

We calculated correlations between the dot memory, span board and composite VSTM scores and total brain volume (TBV), in both bivariate Pearson's correlations and partial correlations controlling for sex. Next, we used partial correlations to assess associations between the dot memory, span board, and composite scores and ROI GMV, controlling for sex. For comparison with previous work, these partial correlations were conducted with and without controlling for TBV. ROI analyses were false discovery rate corrected for multiple comparisons, and statistical significance was set at $p < 0.05$.

## 2.5 Whole-brain statistics

To test our hypothesis that VSTM is related to GMV of specific regions of the visual cortex and right posterior IPS, we elected to utilize an ROI-based approach. However, considering this approach might miss associations in brain areas outside of our hypothesized regions, which could be of interest in comparison to previous work (Konstantinou et al., 2017), we utilized voxel-based morphometry to assess associations between VSTM task performance and GMV throughout the whole-brain. Images were smoothed using an 8mm full-width at half-maximum Gaussian smoothing kernel and entered into whole-brain general linear models, in which task performance on the VSTM tasks were regressed against voxel-wise GMV while controlling for sex, with and without controlling for TBV. Initial voxel-wise threshold was set at $p < 0.001$, followed by family-wise error multiple comparison correction with cluster-wise thresholding at $p < 0.05$.

## 3 Results

## 3.1 VSTM task performance

Raw scores on the span board task ranged from 75 to 138 (mean=111.1, std=18.8), and the dot memory task from 88 to 125 (mean=114.3, std=8.8). The two tasks were significantly correlated (r=0.75, p<0.001).

## 3.2 Association between ROI GMV and VSTM

Before testing individual ROIs, we assessed associations between VSTM scores and TBV. We found that the dot memory task and composite score were significantly correlated with TBV before (r=0.459 and 0.404 respectively, p<0.05) and after (r=0.467 and 0.368 respectively, p<0.05) controlling for sex. For the dot memory task, with exclusion of 3 lower limit outliers this correlation fell below significance in Pearson's (r=0.171, p=0.375) and partial correlations controlling for sex (r=0.179, p=0.362). We calculated associations both with and without controlling for TBV in partial correlations with ROI GMV. Before controlling for TBV, we found a trendline correlation between dot memory task performance and V1 GMV (r=0.374, uncorrected p<0.05), however this fell below trendline significance with exclusion of outliers (r=0.185, p=0.346). No significant associations were found after controlling for TBV, though interestingly, correlation coefficients were substantially reduced for all tasks when TBV was included in the model (Figure 1).

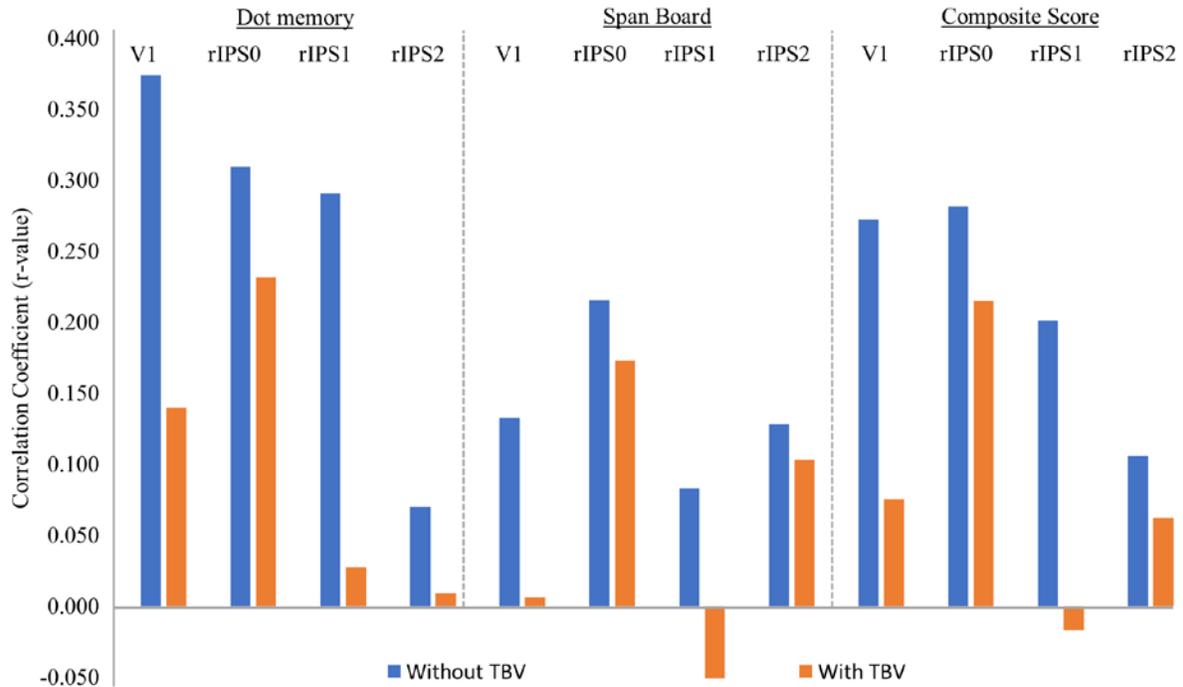

**Figure 1. Regional GMV-VSTM task correlations**. Correlations between gray matter volume and VSTM task (dot memory, span board, and composite score) performance for the four regions of interest are shown before and after correcting for total brain volume. V1 = primary visual cortex; rIPS0-2 = right intraparietal sulcus 0-2; TBV = total brain volume.

**3.3 Association between voxel-wise GMV and VSTM**

In whole-brain voxel-based morphometry analyses without controlling for TBV, we found dot memory scores correlated significantly with two clusters; one encompassing the right basal ganglia, limbic structures, and the hippocampus, with a peak value in the right putamen (MNI x,y,z coordinates: 19.5,9,-4.5; cluster size = 1398 voxels), the other extending from the left parietal lobe towards the superior temporal gyrus, with a peak value in the left Rolandic operculum (MNI x,y,z coordinates: -63,-12,12; cluster size=1168 voxels) (Figure 2). These clusters were no longer significant when TBV was controlled for, or when outliers in the dot memory score were excluded. No significant associations were found with the span board or composite VSTM scores.

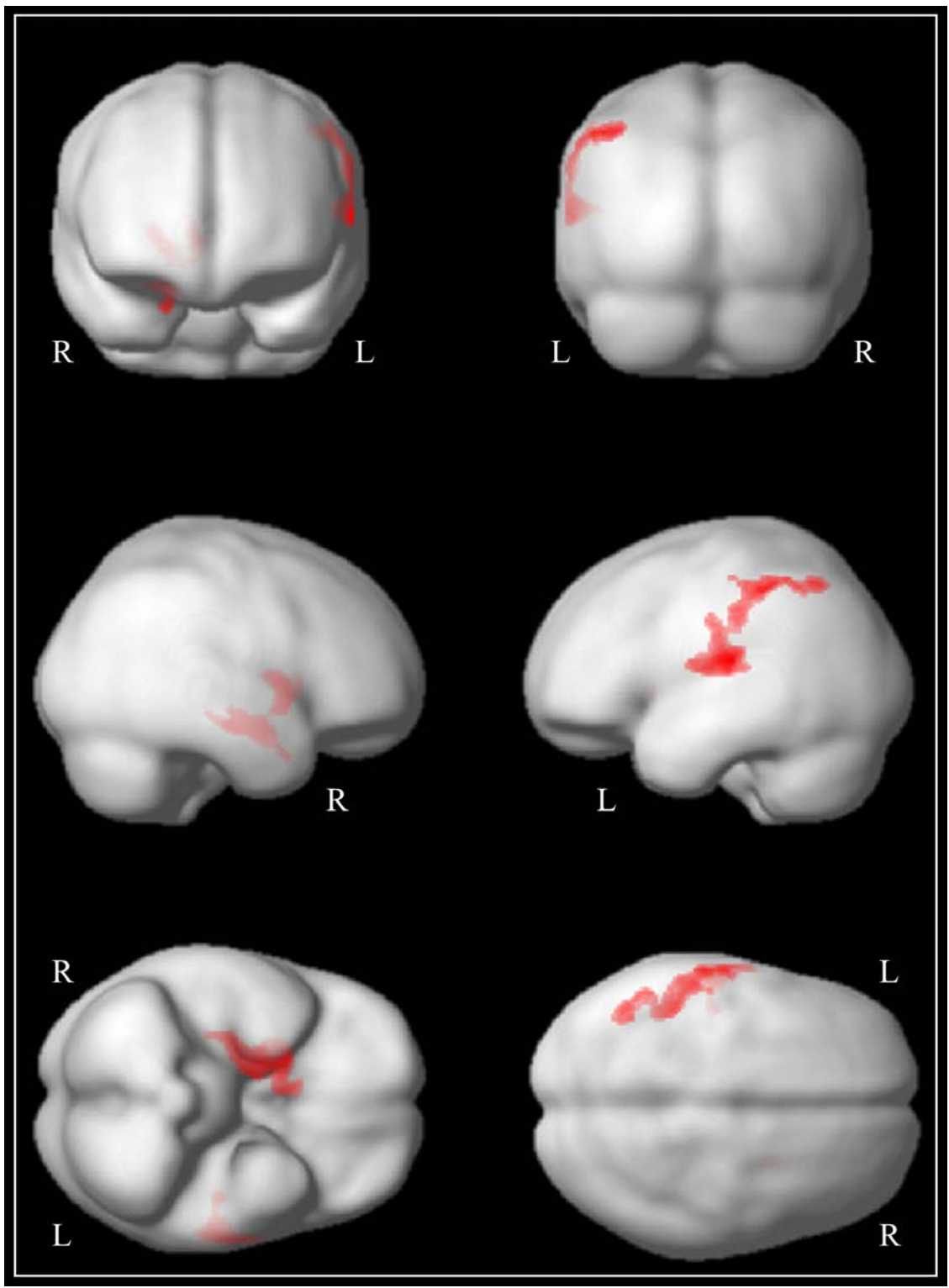

**Figure 2: Assocation between dot memory task and voxel-wise GMV.** Correlations between dot memory task performance and voxel-wise GMV in the whole-brain voxel-based morphometry analysis. Significant clusters are visualized as red overlays on a smoothed 3D image of the MNI template brain (p<0.05).

**4. Discussion**

Motivated by predictions about dorsal stream vulnerability and map theories of VSTM capacity, this study assessed correlations between VSTM performance and GMV in V1 and posterior IPS regions. For comparison with previous studies, we also assessed associations between voxel-wise GMV and VSTM at the whole-brain level. We found that TBV, but not individual regions, was a significant predictor of performance on the dot memory and VSTM composite scores in a sample of healthy young adults. At the whole-brain level, we found that GMV of clusters centered around the right putamen and left Rolandic operculum was associated with dot memory scores, though only when TBV was not accounted for. These results suggest that associations between VSTM and regional GMV are weak when TBV is accounted for.

The findings in this study are in contrast to recent work showing a significant association between V1 volume and VSTM capacity (Bergmann et al., 2016). We note however that this study did not control for TBV in their analyses. Konstantinou et al., (2017) conducted a whole brain analysis to identify neuroanatomical correlates of location VSTM while controlling for TBV. They found that GMV in the left inferior parietal lobule, but not in V1, associated with VSTM. These findings are similar to that of our whole-brain analysis, wherein we found that performance on the dot memory task associated with GMV of left parietal and temporal regions, including a portion of the left inferior parietal lobule. Interestingly however, this association was only significant when TBV was not controlled for. Taken together these two studies and the one reported here suggest that regional GMV-VSTM associations may be relatively weak when TBV is considered, particularly in V1.

It is worth noting that discrepancies between our results and those of previous studies (Bergmann et al., 2016; Konstantinou et al., 2017) could be due to differences in the tasks utilized to assess

VSTM. Konstantinou et al., (2017) showed that object and location VSTM tasks are dependent on the structure of different cortical regions. This might mean that our findings are more comparable to that of Konstantinou et al., (2017), who similarly utilized a location VSTM task, as compared to that of Bergmann et al., (2016), who utilized an object VSTM task. While this might limit the comparability of our results to that of Bergmann et al., (2016), our reported associations between dot memory performance and V1 GMV (r=0.374) suggests that our studies may be similarly sensitive to GMV-VSTM relationships, though we additionally found that the specificity of regional associations was reduced when TBV was accounted for (r=0.185). Another thing to consider in comparing tasks across studies is the extent to which executive functions play a role. It could be argued that tasks involving recollection of both location and order of stimuli, as in our span board task, might recruit cognitive control (Marshuetz & Smith, 2006), making direct comparisons more difficult.

Despite these considerations, we feel that an important contribution of the present work is to demonstrate the importance of considering TBV when calculating brain volume – behavior correlations. We report correlation sizes before and after this correction so that future studies investigating this question may be adequately powered. We further suggest that our data do not necessarily rule out the map account of VSTM capacity (Franconeri et al., 2013), but that perhaps the VSTM-regional GMV relationship is more subtle than previously suggested, or that a distributed set of regions beyond V1 and the posterior IPS may play a role in mediating inter-individual capacity differences.

A limitation of this work is relatively small sample size. Although the study is adequately powered to detect effects of the size reported in previous work (i.e. 86% power to detect r=0.44 anatomical V1 volume association in Bergmann et al.; calculated using gPower), correlations in small samples are prone to inaccurate estimates (Rousselet, basic statistics blog, 2018) and should be interpreted with caution. Considering this, and for the sake of transparency, we report correlations with the dot memory task with and without three lower limit outlier scores included. While removal of these outliers did affect the results, it is important to point out that the three participants in question had a poor performance on both the dot memory and span board tasks, suggesting that these scores accurately reflect the participant's VSTM capacity, and are not due to measurement error.

A second limitation of this work is the use of probabilistic/anatomically defined regions of interest, rather than functionally localized regions. The main reason for this choice is that preliminary analysis of functional localizer data collected on this sample (Bray et al., 2015) showed significant associations between performance on the functional localizer task (essentially a vigilance task) and blood-oxygen-level-dependent (BOLD) response (unpublished). That is, participants who were perhaps more fatigued and therefore less attentive showed lower BOLD amplitude, which could introduce bias. Indeed, studies on visual attention and working memory have shown associations between concurrent performance and  BOLD response (Pessoa et al., 2002; Roalf et al., 2014). We therefore chose to use probabilistic ROIs (Wang et al., 2014), which would no suffer from this issue. However, there is a trade-off, as the mapping to individual neuroanatomy is less accurate. Though Bergmann et al., (2016) found similar results in their study when using functionally vs anatomically defined regions, suggesting this may not have influenced our results.

VSTM is an important cognitive capacity and understanding it's neuroanatomical correlates has both theoretical importance and potential clinical utility. In the present work, we found that posterior IPS regions did not associate with VSTM, and relationships between regional GMV and VSTM was weakened when TBV was controlled for. Our results suggest that regional GMV-VSTM associations, particularly in V1, may be weaker than previously suggested. Moreover, we found a significant association between VSTM and TBV, suggesting VSTM may be more accurately described by distributed representations, in line with findings that VSTM tasks engage distributed cortical regions (e.g. Dotson et al., 2018; Pessoa et al., 2002).

## Declaration of interests

The authors declare that they have no competing interests.

## Funding

This work was supported by an NSERC Postdoctoral Fellowship awarded to SB, an NSERC CGSD award and AIHS Graduate Studentship to DD, and the Hotchkiss Brain Institute. The funding sources were not involved in any aspect of the study, including; study design, data collection, analysis, interpretation of results, writing of the manuscript or decision to publish.

## Acknowledgements

We gratefully acknowledge the participation of all volunteers, and staff at the Seaman Family MR Research Centre; Dr. Mounir Nour and Eileen Pyra.

**References**

Ashburner, J., & Friston, K. J.  (2000) Voxel-based morphometry--the methods *Neuroimage*. Neuroimage, 11(6 Pt 1), 805-21.

Ashburner, J.  (2007) A fast diffeomorphic image registration algorithm *Neuroimage*. Neuroimage, 38(1), 95-113.

Atkinson, J., Braddick, O., Anker, S., Curran, W., Andrew, R., Wattam-Bell, J., & Braddick, F.  (2003) Neurobiological models of visuospatial cognition in children with Williams syndrome: measures of dorsal-stream and frontal function *Dev Neuropsychol*. Dev Neuropsychol, 23(1-2), 139-72.

Attout, L., Noël, M. P., Nassogne, M. C., & Rousselle, L.  (2017) The role of short-term memory and visuo-spatial skills in numerical magnitude processing: Evidence from Turner syndrome P*LoS ONE*. PLoS ONE, 12(2), e0171454.

Baddeley, A.  (2003) Working memory: looking back and looking forward N*at. Rev. Neurosci*. Nat. Rev. Neurosci, 4(10), 829-39.

Bergmann, J., Genç, E., Kohler, A., Singer, W., & Pearson, J.  (2016) Neural Anatomy of Primary Visual Cortex Limits Visual Working Memory Ce*reb. Cortex*. Cereb. Cortex, 26(1), 43-50.

Blair, J. R., & Spreen, O.  (1989) Predicting premorbid IQ: a revision of the National Adult Reading Test Th*e Clinical Neuropsychologist*. The Clinical Neuropsychologist.

Braddick, O., Atkinson, J., & Wattam-Bell, J. (2003) Normal and anomalous development of visual motion processing: motion coherence and 'dorsal-stream vulnerability' Ne*uropsychologia.* Neuropsychologia, 41(13), 1769-84.

Bray, S., Almas, R., Arnold, A. E., Iaria, G., & MacQueen, G. (2015) Intraparietal sulcus activity and functional connectivity supporting spatial working memory manipulation Ce*reb. Cortex.* Cereb. Cortex, 25(5), 1252-64.

Bray, S., Arnold, A. E., Iaria, G., & MacQueen, G. (2013) Structural connectivity of visuotopic intraparietal sulcus Ne*uroimage.* Neuroimage, 82, 137-45.

Bray, S., Dunkin, B., Hong, D. S., & Reiss, A. L. (2011) Reduced functional connectivity during working memory in Turner syndrome Ce*reb. Cortex.* Cereb. Cortex, 21(11), 2471-81.

Buchanan, L., Pavlovic, J., & Rovet, J. (1998) A reexamination of the visuospatial deficit in turner syndrome: De*velopmental Neuropsychology.* Developmental Neuropsychology, 14(2), 341 — 367.

Conway, A. R., Cowan, N., & Bunting, M. F. (2001) The cocktail party phenomenon revisited: the importance of working memory capacity Psyc*hon Bull Rev. Ps*ychon Bull Rev, 8(2), 331-5.

Corbetta, M., & Shulman, G. L. (2002). Control of goal-directed and stimulus-driven attention in the brain. *Nat Rev Neurosci, 3*(3), 201-215

Cornoldi, C., Marconi, F., & Vecchi, T. (2001) Visuospatial working memory in Turner's syndrome *Brain And Cognition. Br*ain And Cognition, 46(1-2), 90-94.

Costanzo, F., Varuzza, C., Menghini, D., Addona, F., Gianesini, T., & Vicari, S. (2013) Executive functions in intellectual disabilities: a comparison between Williams syndrome and Down syndrome Res *Dev Disabil. Re*s Dev Disabil, 34(5), 1770-80.

Cowan, N. (2001) The magical number 4 in short-term memory: a reconsideration of mental storage capacity Beha*v Brain Sci. Be*hav Brain Sci, 24(1), 87-114; discussion 114-85.

Dotson, N. M., Hoffman, S. J., Goodell, B., & Gray, C. M. (2018) Feature-Based Visual Short-Term Memory Is Widely Distributed and Hierarchically Organized Neur*on. Ne*uron.

Franconeri, S. L., Alvarez, G. A., & Cavanagh, P. (2013) Flexible cognitive resources: competitive content maps for attention and memory Tren*ds Cogn. Sci. (Regul. Ed.). Tr*ends Cogn. Sci. (Regul. Ed.), 17(3), 134-41.

Girard, T. A., Wilkins, L. K., Lyons, K. M., Yang, L., & Christensen, B. K. (2018) Traditional test administration and proactive interference undermine visual-spatial working memory performance in schizophrenia-spectrum disorders Cogn *Neuropsychiatry. Co*gn Neuropsychiatry, 1-12.

Greenberg, A. S., Verstynen, T., Chiu, Y. C., Yantis, S., Schneider, W., & Behrmann, M. (2012) Visuotopic cortical connectivity underlying attention revealed with white-matter tractography J. N*eurosci. J.* Neurosci, 32(8), 2773-82.

Konstantinou, N., Constantinidou, F., & Kanai, R. (2017) Discrete capacity limits and neuroanatomical correlates of visual short-term memory for objects and spatial locations Hum *Brain Mapp. Hu*m Brain Mapp, 38(2), 767-778.

Lauritzen, T. Z., D'Esposito, M., Heeger, D. J., & Silver, M. A. (2009). Top-down flow of visual spatial attention signals from parietal to occipital cortex. *J Vis, 9*(13), 18.11-14.

Marshuetz, C., & Smith, E. E. (2006). Working memory for order information: multiple cognitive and neural mechanisms. *Neuroscience, 139*(1), 195-200.

Mazziotta, J. C., Toga, A. W., Evans, A., Fox, P., & Lancaster, J. (1995). A probabilistic atlas of the human brain: theory and rationale for its development. The International Consortium for Brain Mapping (ICBM). *Neuroimage, 2*(2), 89-101.

Miyake, A., Friedman, N. P., Rettinger, D. A., Shah, P., & Hegarty, P. (2001) How are visuospatial working memory, executive functioning, and spatial abilities related? A latent-variable analysis Jour*nal Of Experimental Psychology-general. Jo*urnal Of Experimental Psychology-general, 130(4), 621-640.

Molko, N., Cachia, A., Riviere, D., Mangin, J. F., Bruandet, M., Le Bihan, D., Cohen, L., & Dehaene, S. (2003) Functional and structural alterations of the intraparietal sulcus in a developmental dyscalculia of genetic origin Neur*on. Ne*uron, 40(4), 847-58.

Molko, N., Cachia, A., Rivière, D., Mangin, J. F., Bruandet, M., Le Bihan, D., Cohen, L., & Dehaene, S. (2003) Functional and structural alterations of the intraparietal sulcus in a developmental dyscalculia of genetic origin Neur*on. Neu*ron, 40(4), 847-58.

Nelson, H. E., & Willison, J. (1991) National Adult Reading Test (NART).

Park, D. C., Lautenschlager, G., Hedden, T., Davidson, N. S., Smith, A. D., & Smith, P. K. (2002) Models of visuospatial and verbal memory across the adult life span Psych*ol Aging. Psy*chol Aging, 17(2), 299-320.

Pessoa, L., Gutierrez, E., Bandettini, P., & Ungerleider, L. (2002) Neural correlates of visual working memory: fMRI amplitude predicts task performance Neuro*n. Neu*ron, 35(5), 975-87.

Riggs, K. J., Simpson, A., & Potts, T. (2011) The development of visual short-term memory for multifeature items during middle childhood J Exp *Child Psychol. J E*xp Child Psychol, 108(4), 802-9.

Roalf, D. R., Ruparel, K., Gur, R. E., Bilker, W., Gerraty, R., Elliott, M. A., Gallagher, R. S., Almasy, L., Pogue-Geile, M. F., Prasad, K., Wood, J., Nimgaonkar, V. L., & Nimgaonkar, V. L. (2014) Neuroimaging predictors of cognitive performance across a standardized neurocognitive battery Neuro*psychology. Neu*ropsychology, 28(2), 161-76.

Rousselet G. (2018) basic statistics blog: Small n correlations cannot be trusted; https://garstats.wordpress.com/2018/06/01/smallncorr/; Accessed July 11 2018.

Sereno, M. I., Dale, A. M., Reppas, J. B., Kwong, K. K., Belliveau, J. W., Brady, T. J., Rosen, B. R., & Tootell, R. B. (1995) Borders of multiple visual areas in humans revealed by functional magnetic resonance imaging Scien*ce. Sci*ence, 268(5212), 889-93.

Silver, M. A., & Kastner, S. (2009) Topographic maps in human frontal and parietal cortex Trend*s Cogn. Sci. (Regul. Ed.). Tre*nds Cogn. Sci. (Regul. Ed.), 13(11), 488-95.

Vogel, E. K., & Machizawa, M. G. (2004) Neural activity predicts individual differences in visual working memory capacity Natur*e. Nat*ure, 428(6984), 748-751.

Vogel, E. K., McCollough, A. W., & Machizawa, M. G. (2005) Neural measures reveal individual differences in controlling access to working memory Natur*e. Nat*ure, 438(7067), 500-3.

Wang, L., Mruczek, R. E., Arcaro, M. J., & Kastner, S.  (2014) Probabilistic Maps of Visual Topography in Human Cortex Cereb. *Cortex. Cer*eb. Cortex.

Westerberg, H., Hirvikoski, T., Forssberg, H., & Klingberg, T.  (2004) Visuo-spatial working memory span: A sensitive measure of cognitive deficits in children with ADHD Child *Neuropsychology. Chi*ld Neuropsychology, 10(3), 155-161.

Zhao, C., & Gong, G.  (2017) Mapping the effect of the X chromosome on the human brain: Neuroimaging evidence from Turner syndrome Neuro*sci Biobehav Rev. Neu*rosci Biobehav Rev, 80, 263-275.